\documentclass[twocolumn]{aastex631}

\turnoffediting

\hypersetup{
   colorlinks,
   linkcolor={blue!88!black!80},
   citecolor={blue!88!black!80},
   urlcolor={blue!88!black!80}}

\usepackage{gensymb}
\usepackage{graphicx}

\newcommand{\ha}{H$\alpha$}

\newcommand{\brg}{Br$\gamma$}
\newcommand{\pfa}{Pf$\alpha$}
\newcommand{\oiv}{[\ion{O}{4}]}
\newcommand{\oiii}{[\ion{O}{3}]}

\newcommand{\arii}{[\ion{Ar}{2}]}
\newcommand{\feii}{[\ion{Fe}{2}]}
\newcommand{\feviii}{[\ion{Fe}{8}]}

\newcommand{\ci}{[\ion{C}{1}]}
\newcommand{\nii}{[\ion{N}{2}]}
\newcommand{\mgvii}{[\ion{Mg}{7}]}

\newcommand{\mgv}{[\ion{Mg}{5}]}

\newcommand{\sivi}{[\ion{Si}{6}]}
\newcommand{\nevi}{[\ion{Ne}{6}]}
\newcommand{\molhy}{H$_2$}
\newcommand{\um}{$\mu$m}

\received{September 2, 2022}
\revised{September 27, 2022}
\accepted{\today}

\submitjournal{ApJL}

\shorttitle{Circumnuclear Gas in NGC 7469}
\shortauthors{U et al.}
\graphicspath{{./}{figures/}}

\begin{document}

\title{GOALS-JWST: Resolving the Circumnuclear Gas Dynamics in NGC 7469 in the Mid-Infrared}

\correspondingauthor{Vivian U}
\email{vivianu@uci.edu}

\author[0000-0002-1912-0024]{Vivian U}
\affiliation{Department of Physics and Astronomy, 4129 Frederick Reines Hall, University of California, Irvine, CA 92697, USA}

\author[0000-0001-8490-6632]{Thomas Lai}
\affiliation{IPAC, California Institute of Technology, 1200 E. California Blvd., Pasadena, CA 91125}

\author[0000-0002-6570-9446]{Marina Bianchin}
\affiliation{Universidade Federal de Santa Maria, Departamento de Física, Centro de Ciências Naturais e Exatas, 97105-900, Santa Maria, RS, Brazil}
\affiliation{Department of Physics and Astronomy, 4129 Frederick Reines Hall, University of California, Irvine, CA 92697, USA}

\author[0000-0002-0164-8795]{Raymond P. Remigio}
\affiliation{Department of Physics and Astronomy, 4129 Frederick Reines Hall, University of California, Irvine, CA 92697, USA}

\author[0000-0003-3498-2973]{Lee Armus}
\affiliation{IPAC, California Institute of Technology, 1200 E. California Blvd., Pasadena, CA 91125}

\author[0000-0003-3917-6460]{Kirsten L. Larson}
\affiliation{AURA for the European Space Agency (ESA), Space Telescope Science Institute, 3700 San Martin Drive, Baltimore, MD 21218, USA}

\author[0000-0003-0699-6083]{Tanio D\'iaz-Santos}
\affiliation{Institute of Astrophysics, Foundation for Research and Technology-Hellas (FORTH), Heraklion, 70013, Greece}
\affiliation{School of Sciences, European University Cyprus, Diogenes street, Engomi, 1516 Nicosia, Cyprus}

\author[0000-0003-2638-1334]{Aaron Evans}
\affiliation{National Radio Astronomy Observatory, 520 Edgemont Rd, Charlottesville, VA, 22903, USA}
\affiliation{Department of Astronomy, University of Virginia, 530 McCormick Road, Charlottesville, VA 22903, USA}

\author[0000-0002-2596-8531]{Sabrina Stierwalt}
\affiliation{Physics Department, 1600 Campus Road, Occidental College, Los Angeles, CA 90041, USA}

\author[0000-0002-9402-186X]{David R.~Law}
\affiliation{Space Telescope Science Institute, 3700 San Martin Drive, Baltimore, MD 21218, USA}

\author[0000-0001-6919-1237]{Matthew A. Malkan}
\affiliation{Department of Physics \& Astronomy, 430 Portola Plaza, University of California, Los Angeles, CA 90095, USA}

\author[0000-0002-1000-6081]{Sean Linden}
\affiliation{Department of Astronomy, University of Massachusetts at Amherst, Amherst, MA 01003, USA}

\author[0000-0002-3139-3041]{Yiqing Song}
\affiliation{Department of Astronomy, University of Virginia, 530 McCormick Road, Charlottesville, VA 22903, USA}
\affiliation{National Radio Astronomy Observatory, 520 Edgemont Rd, Charlottesville, VA, 22903, USA}
\affiliation{European Southern Observatory, Alonso de Cordova 3107, Santiago, Chile}

\author[0000-0001-5434-5942]{Paul P. van der Werf}
\affiliation{Leiden Observatory, Leiden University, PO Box 9513, 2300 RA Leiden, The Netherlands}

\author[0000-0002-1158-6372]{Tianmu Gao}
\affiliation{Department of Astronomy, Beijing Normal University, Beijing 100875, China}

\author[0000-0003-3474-1125]{George C. Privon}
\affiliation{National Radio Astronomy Observatory, 520 Edgemont Rd, Charlottesville, VA, 22903, USA}
\affiliation{Department of Astronomy, University of Florida, P.O. Box 112055, Gainesville, FL 32611, USA}

\author[0000-0001-7421-2944]{Anne M. Medling}
\affiliation{Department of Physics \& Astronomy and Ritter Astrophysical Research Center, University of Toledo, Toledo, OH 43606,USA}
\affiliation{ARC Centre of Excellence for All Sky Astrophysics in 3 Dimensions (ASTRO 3D); Australia}

\author[0000-0003-0057-8892]{Loreto Barcos-Mu\~noz}
\affiliation{National Radio Astronomy Observatory, 520 Edgemont Rd, Charlottesville, VA, 22903, USA}
\affiliation{Department of Astronomy, University of Virginia, 530 McCormick Road, Charlottesville, VA 22903, USA}

\author[0000-0003-4073-3236]{Christopher C. Hayward}
\affiliation{Center for Computational Astrophysics, Flatiron Institute, 162 Fifth Avenue, New York, NY 10010, USA}

\author[0000-0003-4268-0393]{Hanae Inami}
\affiliation{Hiroshima Astrophysical Science Center, Hiroshima University, 1-3-1 Kagamiyama, Higashi-Hiroshima, Hiroshima 739-8526, Japan}

\author[0000-0002-5807-5078]{Jeff Rich}
\affiliation{The Observatories of the Carnegie Institution for Science, 813 Santa Barbara Street, Pasadena, CA 91101}

\author[0000-0002-5828-7660]{Susanne Aalto}
\affiliation{Department of Space, Earth and Environment, Chalmers University of Technology, 412 96 Gothenburg, Sweden}

\author[0000-0002-7607-8766]{Philip Appleton}
\affiliation{IPAC, California Institute of Technology, 1200 E. California Blvd., Pasadena, CA 91125}

\author[0000-0002-4375-254X]{Thomas Bohn}
\affiliation{Hiroshima Astrophysical Science Center, Hiroshima University, 1-3-1 Kagamiyama, Higashi-Hiroshima, Hiroshima 739-8526, Japan}

\author[0000-0002-5666-7782]{Torsten B\"oker}
\affiliation{European Space Agency, Space Telescope Science Institute, Baltimore, MD 21218, USA}

\author[0000-0002-1207-9137]{Michael J. I. Brown}
\affiliation{School of Physics and Astronomy, Monash University, Clayton, VIC 3800, Australia}

\author[0000-0002-2688-1956]{Vassilis Charmandaris}
\affiliation{Department of Physics, University of Crete, Heraklion, 71003, Greece}
\affiliation{Institute of Astrophysics, Foundation for Research and Technology-Hellas (FORTH), Heraklion, 70013, Greece}
\affiliation{School of Sciences, European University Cyprus, Diogenes street, Engomi, 1516 Nicosia, Cyprus}

\author[0000-0002-1392-0768]{Luke Finnerty}
\affiliation{Department of Physics \& Astronomy, 430 Portola Plaza, University of California, Los Angeles, CA 90095, USA}

\author[0000-0001-6028-8059]{Justin Howell}
\affiliation{IPAC, California Institute of Technology, 1200 E. California Blvd., Pasadena, CA 91125}

\author[0000-0002-4923-3281]{Kazushi Iwasawa}
\affiliation{Institut de Ci\`encies del Cosmos (ICCUB), Universitat de Barcelona (IEEC-UB), Mart\'i i Franqu\`es, 1, 08028 Barcelona, Spain}
\affiliation{ICREA, Pg. Llu\'is Companys 23, 08010 Barcelona, Spain}

\author[0000-0003-2743-8240]{Francisca Kemper}
\affiliation{Institut de Ciencies de l'Espai (ICE, CSIC), Can Magrans, s/n, 08193 Bellaterra, Barcelona, Spain}
\affiliation{ICREA, Pg. Lluís Companys 23, Barcelona, Spain}
\affiliation{Institut d'Estudis Espacials de Catalunya (IEEC), E-08034 Barcelona, Spain}

\author[0000-0001-7712-8465]{Jason Marshall}
\affiliation{4Glendale Community College, 1500 N. Verdugo Rd., Glendale, CA 91208}

\author[0000-0002-8204-8619]{Joseph M. Mazzarella}
\affiliation{IPAC, California Institute of Technology, 1200 E. California Blvd., Pasadena, CA 91125}

\author[0000-0002-6149-8178]{Jed McKinney} 
\affiliation{Department of Astronomy, University of Massachusetts, Amherst, MA 01003, USA.}
\affiliation{Department of Astronomy, The University of Texas at Austin, 2515 Speedway Blvd Stop C1400, Austin, TX 78712, USA}

\author[0000-0002-2713-0628]{Francisco Muller-Sanchez}
\affiliation{Department of Physics and Materials Science, The University of Memphis, 3720 Alumni Avenue, Memphis, TN 38152, USA}

\author[0000-0001-7089-7325]{Eric J.\,Murphy}
\affiliation{National Radio Astronomy Observatory, 520 Edgemont Road, Charlottesville, VA 22903, USA}

\author[0000-0002-1233-9998]{David Sanders}
\affiliation{Institute for Astronomy, University of Hawaii, 2680 Woodlawn Drive, Honolulu, HI 96822}

\author[0000-0001-7291-0087]{Jason Surace}
\affiliation{IPAC, California Institute of Technology, 1200 E. California Blvd., Pasadena, CA 91125}




\begin{abstract}

The nearby, luminous infrared galaxy (LIRG) NGC 7469 hosts a Seyfert nucleus with a circumnuclear star-forming ring and is thus the ideal local laboratory for investigating the starburst--AGN connection in detail. 
We present integral-field observations of the central 1.3 kpc region in NGC 7469 obtained with the \emph{JWST} Mid-InfraRed Instrument. 
Molecular and ionized gas distributions and kinematics at a resolution of $\sim$100 pc over the $4.9-7.6\mu$m region are examined to study gas dynamics influenced by the central AGN. 
The low-ionization \feii~$\lambda$5.34\um~and \arii~$\lambda$6.99\um~lines are bright on the nucleus and in the starburst ring, as opposed to \molhy~S(5) $\lambda$6.91\um~which is strongly peaked at the center and surrounding ISM. The high-ionization \mgv~line is resolved and shows a broad, blueshifted component associated with the outflow.  It has a nearly face-on geometry that is strongly peaked on the nucleus, where it reaches a maximum velocity of $-$650 km s$^{-1}$, and extends about 400 pc to the East. 
Regions of enhanced velocity dispersion in \molhy~and \feii~$\sim$180 pc from the AGN that also show high $L(\mathrm{H}_2)/L(\mathrm{PAH})$ and $L$(\feii)/$L$(\pfa) ratios to the W and N of the nucleus pinpoint regions where the ionized outflow is depositing energy, via shocks, into the dense interstellar medium between the nucleus and the starburst ring.
These resolved mid-infrared observations of the nuclear gas dynamics demonstrate the power of \emph{JWST} and its high-sensitivity integral-field spectroscopic capability to resolve feedback processes around supermassive black holes in the dusty cores of nearby LIRGs. 

\end{abstract}



\section{Introduction} \label{sec:intro}
Accreting supermassive black holes (SMBHs) within active galactic nuclei (AGNs) are thought to play a prominent role in influencing the interstellar medium (ISM) of their host through feedback mechanisms such as outflows~\cite[see reviews by][]{Veilleux20,Armus20}. Distinguishing the drivers of such winds and determining their direct impacts on surrounding star formation has been difficult due to observational challenges such as heavy dust obscuration. Investigating the triggering of AGN outflows is most ideally done close to the launching site which is often obscured by dust in the case of many Seyfert and luminous infrared galaxies (LIRGs; $L_\mathrm{IR} \geq 10^{11} L_\odot$), with the latter exhibiting extinction upward of $A_V \sim 40$ mag or higher~\citep{Mattila07,Vaisanen17,U19,Falstad21,Perez-Torres21}. 

NGC 7469 offers a prime opportunity to study these phenomena in detail. It is a nearby ($z = 0.01627$; $D_L =$ 70.6 Mpc)\footnote{NED}
LIRG hosting a bright Seyfert 1.5 nucleus~\citep{Landt08} surrounded by a starburst ring~\cite[$R_\mathrm{peak} \sim$ 530 pc;][]{Genzel95,Song21} with a bimodal age distribution of stellar populations~\citep{Diaz-Santos07}. 
Evidence of a wide-angle E-W \textrm{biconically illuminated} outflow in the near-infrared coronal line \sivi~$\lambda$1.96\um~(R$_\mathrm{[Si VI]}$ = 90 pc) has been presented using VLT/SINFONI 
integral field spectroscopy (IFS)~\cite[FWHM = 0\farcs14, or 46 pc;][]{Muller-Sanchez11}. The characteristic outflow structure with broad and blueshifted components along the minor axis \textrm{of the galactic disk} was not seen in the coarser optical GTC/MEGARA IFS observations~\cite[FWHM = 0\farcs93, or 307 pc;][]{Cazzoli20}. However, turbulent, non-circular kinematics, detected in a component of the \ha-\nii~complex within the central 610 pc, might be associated with it. More recently, extended circumnuclear outflows were found reaching 531 pc from the AGN in VLT/MUSE observations of \oiii~$\lambda$5007\AA~\cite[FWHM = 1\farcs23, or 406 pc;][]{Robleto-Orus21,Xu22}. 
ALMA observations of CO($1-0$), CO($2-1$), and \ci~in the inner $\sim$ 2 kpc region with angular resolutions of 0\farcs35 (120 pc) show largely rotational kinematics
\citep{Izumi20,Nguyen21}. Despite the multiwavelength effort, there is still no clear picture for how the putative outflow interacts with the circumnuclear gas and starburst ring.

With \emph{JWST}'s 6.5-meter mirror and advanced instrument suite, we can now resolve the dynamics of gas and dust at angular resolution of $\sim$ 0\farcs2$-$0\farcs8 in the mid-infrared wavelengths (5$-$28\um). In particular, this \emph{Letter} investigates the detailed gas kinematics in the inner 600-pc region (the ``inner ISM region" hereafter) of NGC 7469. We first describe the observations and data processing in Section \ref{sec:data}. Section \ref{sec:analysis} highlights the moment maps of several key ionized and molecular gas emission lines and demonstrates the richness of features in the mid-infrared spectra extracted from the inner ISM region. In Section \ref{sec:discussion}, we 
discuss the properties of the detected outflow and the shock excitation in the ISM. In several companion papers, we investigate the infrared spectral properties of the AGN (Armus et al. 2022, in preparation), the starburst ring~\citep{Lai22}, and the circumnuclear star-forming regions~\citep{Bohn22} in NGC 7469.

Throughout the paper, $H_0 = 70$\,km\,s$^{-1}$\,Mpc$^{-1}$, $\Omega_{\rm m}$ = 0.30, and $\Omega_{\rm vac}$ = 0.70 have been adopted. At the redshift of $z = 0.01627$, 1\arcsec = 330 pc~\citep{Wright06}.

\section{Observations and Data Reduction} \label{sec:data}
Mid-infrared IFS observations of NGC 7469 were taken with the \emph{JWST} Mid-InfraRed Instrument~\cite[MIRI;][]{Rieke15,Labiano21} in Medium Resolution Spectroscopy (MRS) mode on 2022 July 3-4 UT as part of the Early Release Science (ERS) Program 1328 (Co-PIs: L. Armus and A. Evans). The observations covered the full 4.9$-$28.8\um~range using the short (A), medium (B), and long (C) sub-bands in all four channels. We adopted the FASTR1 readout pattern to optimize the dynamic range expected in the observations. Using 
the extended source 4-pt dither pattern,
the science exposure time per sub-band was 444 seconds. Because our source is extended, we linked the observation to a dedicated background with the same observational parameters in all three grating settings. 

Uncalibrated science and background observations were downloaded using the MAST Portal and processed with the \emph{JWST} Science Calibration Pipeline \citep{jwstpipe} version 1.6$+$ in batch mode. The \texttt{Detector1} pipeline applies detector-level corrections and ramp fitting to the individual exposures. The output rate images were subsequently processed outside the \emph{JWST} pipeline to flag newly-acquired bad pixels and additional cosmic ray artifacts, and to remove vertical stripes and zeropoint residuals remaining after the pipeline dark subtraction.  These additional corrections broadly follow the steps taken for \emph{JWST} ERO observations as described by \citet{pontop22}.  The resulting rate files are then processed with the \emph{JWST} \texttt{Spec2} pipeline for distortion and wavelength calibration, flux calibration, and other 2d detector level steps.
Residual fringe corrections using prototype pipeline code have been applied to both the Stage 2 products and to the 1-D spectra resulting from Stage 3 processing.

Stage 3 processing (\texttt{Spec3}) performs background subtraction before combining data from multiple exposures into the final data cubes. Background light is subtracted from the 2-D science images using a master background frame generated from our associated background observations. The master background is a 1-D median sigma-clipped spectrum calculated over the field-of-view (FOV) of the background observations and projected to the entire 2-D detector array. Since it is a combination of many detector pixels, it does not degrade the signal-to-noise ratio (SNR) the way a pixel-by-pixel background subtraction would.
The cube-building step in \texttt{Spec3} assembles a single 3-D data cube from all of the individual 2-D calibrated detector images, combining data across individual wavelength bands and channels.

NIRCam's broadband F150W, F200W, and F444W images of NGC 7469 from our ERS-1328 Program have been included in this paper for visualization purposes. Readers are referred to~\citet{Bohn22} for the data processing details.
 
\section{Analysis and Results} \label{sec:analysis}

\subsection{Emission Line Maps}
In this paper, we focus our investigation on the Channel 1 MRS data (4.89\um~to 7.66\um), which hosts key diagnostic lines for shock excitation (\feii~and \molhy), strength of the radiation field (\arii), and coronal region (\mgv) at the highest spatial resolution afforded by MRS at 0\farcs3. 
Known issues with MRS wavelength solutions (at the time of this writing) are generally not a concern for Ch 1, where zeropoint variations at the 0.001\um~level ($\sim$ 60 km s$^{-1}$) are within the spectral noise at these wavelengths. 
We fit single Gaussian profiles to three of the brightest ionized and molecular lines in Ch 1 (\feii~$\lambda$5.34\um, \molhy~S(5)~$\lambda$6.91\um, and \arii~$\lambda$6.99\um) on a spaxel-by-spaxel basis and computed their corresponding flux, velocity, and velocity dispersion maps using \texttt{ifscube}~\citep{Ruschel-Dutra20,Ruschel-Dutra21}. 
We correct our linewidth measurements for instrumental broadening ($\sigma_\mathrm{inst}$ = 36.5 km s$^{-1}$) by subtracting it in quadrature.

\begin{figure*}[!bhtp]
  \includegraphics[width=\textwidth]{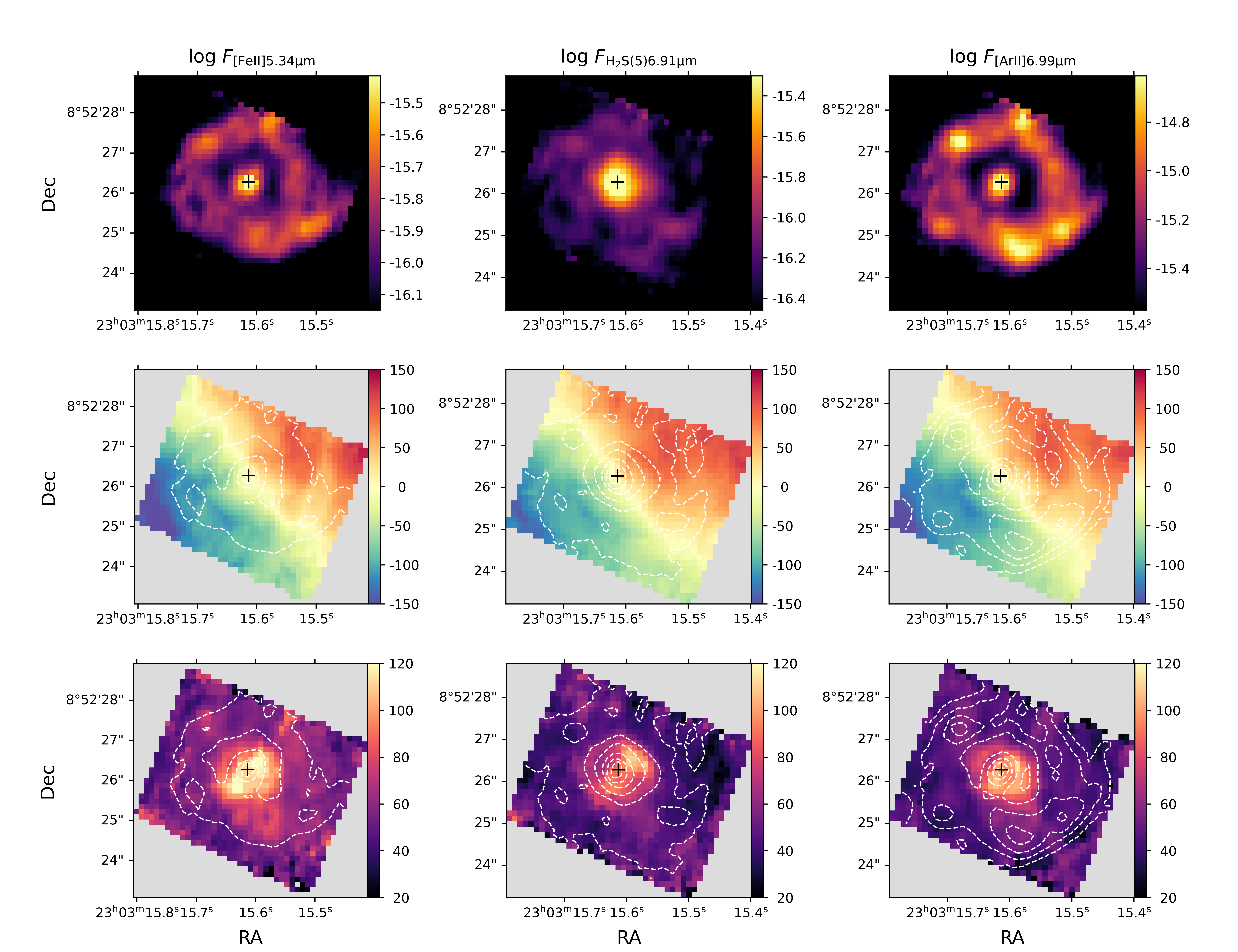}
  \caption{The distribution of flux (top; in log erg s$^{-1}$ cm$^{-2}$ pixel$^{-1}$), velocity (middle; in km s$^{-1}$), and velocity dispersion (bottom; in km s$^{-1}$) for several bright emission lines in Ch1: \feii~$\lambda$5.34\um~(left), \molhy~S(5)~$\lambda$6.91\um~(middle), \arii~$\lambda$6.99\um~(right). Contours based on the flux distribution are overlaid on the kinematic maps of the respective gas species. The black cross marks the peak of the \molhy~emission. 
  North is up and East is to the left. 
  The distributions of the molecular and ionized gas emission are strikingly different, with \molhy~being preferentially bright at the center while \feii~and \arii~appear clumpy at the star-forming ring. 
\molhy~and \feii~exhibit enhanced dispersion $\sim$ 180 pc off the nucleus.
  \label{fig:linemaps}}
\end{figure*}

As shown in 
Figure~\ref{fig:linemaps}, the morphologies of the emission lines differ among these molecular and low-ionization potential (IP) gases: \feii~(IP = 7.9 eV) and \arii~(IP = 15.8 eV) are bright and compact at the nucleus and at several star-forming clumps in the ring. 
The starburst ring appears slightly asymmetric around the central nucleus, where the southeast (SE) inner edge is closer to the AGN. 
In contrast, the \molhy~dominates and appears extended at the center but relatively weak in the ring. The nuclear \molhy~emission is $\sim$1\farcs4 across, nearly 5 times larger than the FWHM of the point spread function (PSF) of 0\farcs3 at 7.2\um~continuum. Filamentary structures extend in the NE and SW directions from the central emission that appear to be well aligned with high surface brightness CO gas~\cite[see Figure 2 in][]{Izumi20} at the sites of the innermost spiral arms.

The gas kinematics for these lines are largely similar, sharing a bulk rotational motion around the ring. The kinematic major axis has a position angle (PA) of 126\textdegree~as measured using \texttt{PaFit}~\citep{Krajnovic06}. 
\textrm{The morphological semi-major axis (2\farcs1) and semi-minor axis (1\farcs7) of the ring gives an inclination angle $i =$ 51\textdegree.}
The velocity field is reminiscent of a tilted rotating ring with the \textrm{near (far)} side in the E/NE (W/SW). 
We compute the contribution to the velocity dispersion due to rotation and beam smearing following the technique from~\citet{Swinbank12,DeBreuck14}: we measure the luminosity-weighted velocity gradient across the FWHM of the beam at each spaxel and subtracted it from the velocity dispersion in quadrature. 
For all these gases, the median effect was at the 6\% level ($\sim$ 4 km s$^{-1}$), with 75\% of the spaxels affected under 10\% ($\sim$ 7 km s$^{-1}$) and 95\% under 20\% ($\sim$ 14 km s$^{-1}$).

The \molhy~velocity dispersion is slightly suppressed at the center \textrm{(95 km s$^{-1}$, or 75 km s$^{-1}$ intrinsically)} but increases toward the NW direction in a cone shape, similar to the shocked \molhy~$\lambda$2.12\um~detected in Mrk 273~\citep{U13}. The \feii~gas, sensitive to strong shocks, shows high velocity dispersions \textrm{(reaching as high as 135 km s$^{-1}$)} roughly along the kinematic axis toward both NW and SE. 

\subsection{Spectral Fits}

\begin{figure*}[hbt!]
  \includegraphics[trim=0 49 0 0,clip,width=0.44\textwidth]{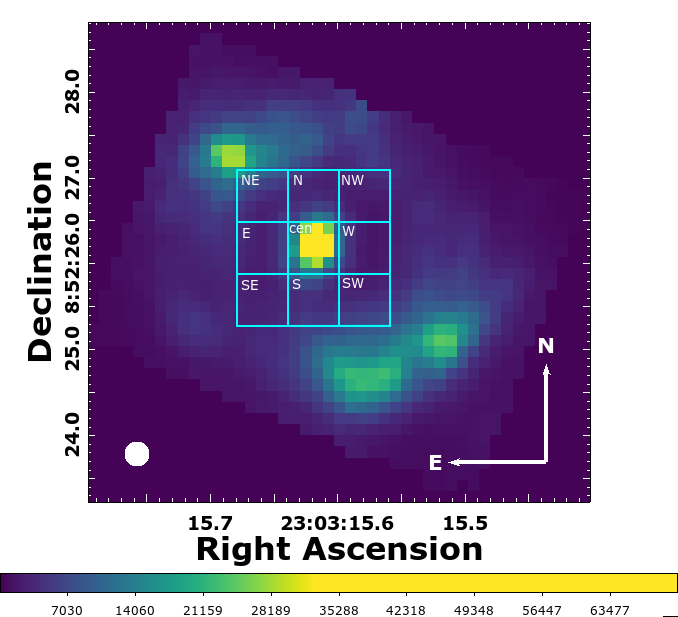} 
  \includegraphics[width=0.57\textwidth]{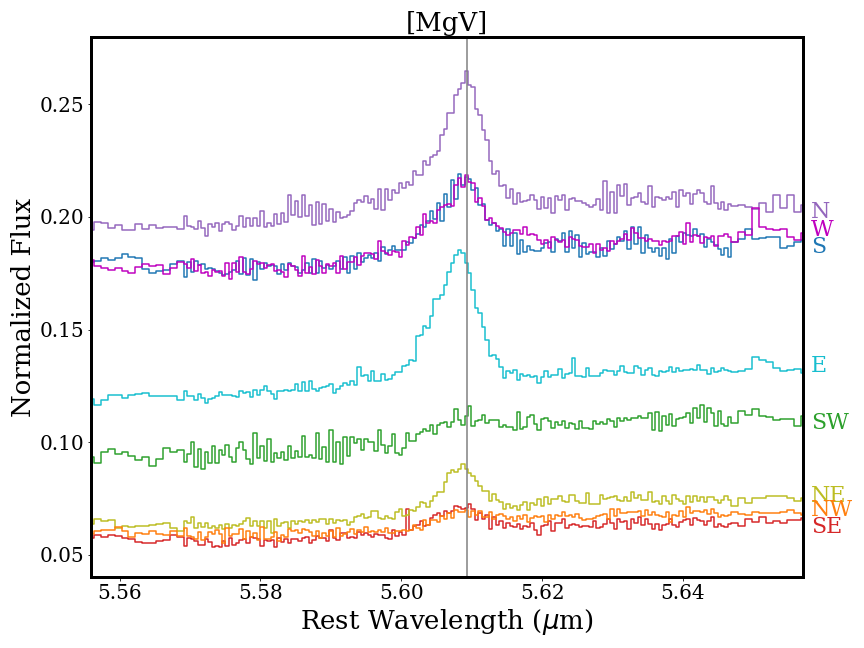}
  \includegraphics[width=\textwidth]{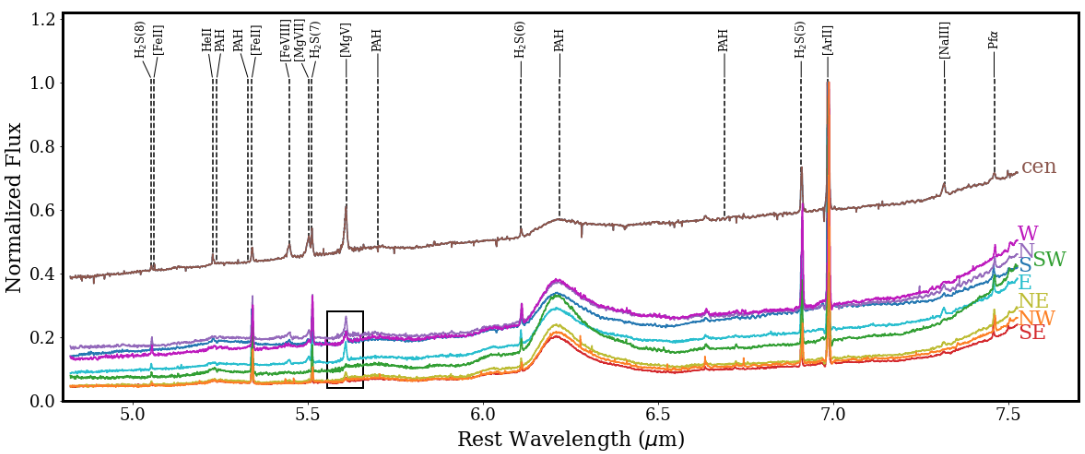}
  \begin{picture}(0,0)
    \put(153,76){\rotatebox{58.5}{\rlap{\makebox[7.2cm]{\dotfill}}}}
    \put(170,76){\rotatebox{27.5}{\rlap{\makebox[13.1cm]{\dotfill}}}}
  \end{picture}
  \caption{(Top left) Extraction grid (cyan) overlaid on the Channel 1 flux image at 7.1\um. Each extraction region is 0\farcs6 on a side, capturing the inner ISM region in this 3$\times$3 grid. 
  The PSF FWHM is marked with a filled circle in the bottom left corner. (Bottom) The Ch 1 short-medium-long stitched spectra extracted from the grid, each labeled by its direction relative to the center `cen'. Spectral features are labeled; the spectra are normalized at \arii~$\lambda$6.99\um~for ease of comparison. A box around coronal line \mgv~$\lambda$5.61\um~is drawn. (Inset) A zoomed-in view of the spectral region centered on \mgv. 
  A vertical line at 5.609\um~as defined based on the peak of \mgv~at the central grid indicates the systemic velocity. \mgv~is present in most of the inner ISM regions (except for NW and SW); its blueshifted flux is most pronounced in the E. 
  \label{fig:specs}} 
\end{figure*}

To quantify the ISM conditions surrounding the AGN, we divide the inner 1\farcs8 (600 pc) ISM region within the ring as seen at 7.1\um, the observed wavelength of \arii, into a 3 $\times$ 3 grid and subsequently extract 1-D spectra from these bins.
Since the central region contains a point source from the AGN, we apply a wavelength-dependent aperture correction to the central spectrum. The aperture correction is calculated using the MRS PSF models from WebbPSF~\citep{Perrin14} that have been adjusted to match preliminary in-flight performance during \emph{JWST} commissioning (private communication). These models are nearly diffraction limited longward of 8\um~and moderately elliptical at shorter wavelengths.
Figure~\ref{fig:specs} shows the extraction regions (0\farcs6 on a side) and their corresponding spectra normalized at \arii. 
The stitched Ch 1 short-medium-long spectra exhibit a number of strong molecular and fine structure lines, as well as polycyclic aromatic hydrocarbon (PAH) features. Several of these emission lines were previously seen with \emph{Spitzer}/IRS, but with a large aperture that included the starburst ring and the AGN~\citep{Stierwalt13,Inami13}. The relatively poor spatial resolution of \emph{Spitzer} precluded studies of the variations in the strengths and profile shapes of these lines with positions on these scales, now made possible with \emph{JWST}. Detailed discussions of the nuclear AGN spectrum and of the dust grains in the star-forming ring are presented in two companion papers~\cite[Armus et al. 2022, in preparation,][respectively]{Lai22}. 

A handful of high-ionization lines (\feviii~$\lambda$5.48\um~with IP = 124 eV, \mgvii~$\lambda$5.50\um~with IP = 186 eV, and \mgv~$\lambda$5.61\um~with IP = 109 eV) show clear blue-shifted wings. The coronal line \mgv~is the strongest among these features in all the 1-D spectra. Given their high IPs, photo-ionized coronal lines are 
typically produced on several hundred parsecs away from the AGN~\citep{Ferguson97}. Kinematics of coronal lines often reveal blueshifts indicating that outflows on these scales are common~\cite[e.g.~][]{Riffel21}.
While strong coronal lines might be expected at the nucleus, broad and blueshifted components may be present in the immediate vicinity of the nucleus if an outflow is present.
Indeed, a close look at our extracted spectra reveals that \mgv~is most significantly shifted in the E, 70 km s$^{-1}$ with respect to the center (Figure~\ref{fig:specs} inset). The blueshifted asymmetric profile of \mgv~can be seen in several other directions as well. 

To further investigate the distribution and kinematics of \mgv, we fit two Gaussian components using a modified version of the Bayesian AGN Decomposition Analysis for SDSS Spectra software package~\cite[\texttt{BADASS};][]{Sexton21} spaxel-by-spaxel and generate moment maps for the two components; see Appendix \ref{sec:mgv}. The core, narrow component of \mgv~is plausibly consistent with rotation at the same PA as that for the low-ionization and molecular gases. A broad component is identified in a subset of the central spaxels, encompassing the blueshifted outflowing gas.
The outflow is detected up to a projected distance of $\sim$420 pc E of the nucleus. Within this physical extent, the gas exhibits a median line-of-sight velocity of $v_\mathrm{med} = -205$ km s$^{-1}$
but reaching as high as $-650$ km s$^{-1}$ close to the AGN.
This outflow is also identified in other high-ionization lines such as \feviii~$\lambda$5.45\um~(IP = 124 eV) and \nevi~$\lambda$7.65\um~(IP = 126 eV), 
and mid-ionization lines such as \oiv~$\lambda$25.9\um~(IP = 55 eV), 
the detailed line profiles for which will be characterized in Armus et al. 2022, in preparation. We do find that the high ionization lines in the nuclear spectrum exhibit extreme blueshifted velocities upward of 1000 km s$^{-1}$, \textrm{indicating the presence of very fast winds at the center.}

Multi-component spectral fitting is performed on the full Ch 1 spectral coverage of each 1-D spectrum using the Continuum And Feature Extraction~\cite[\texttt{CAFE};][D\`iaz-Santos et al. 2022, in preparation]{Marshall07} software package originally written in IDL for analyzing \emph{Spitzer}/IRS data.  While individual atomic or molecular lines can be fit using single or double Gaussian profiles, \texttt{CAFE} specializes in decomposing the emission in the mid-infrared regime into AGN, PAHs, dust of different temperatures, and starburst components, and is ideal for recovering properties of PAH features in 1-D spectra that often require multiple components to fit correctly. The development of an improved version of the \texttt{CAFE} code that handles high-resolution spectral decomposition is fully described in D\`iaz Santos et al. 2022, in preparation). Major updates include a fully-functional Python version applicable to the data format and spectral resolution of \emph{JWST} IFS data.
The flux densities for the various relevant emission-line features are reported in Table \ref{tbl:h2pah}. 

\begin{deluxetable*}{lcccccc}[htb]
\tablecaption{Emission Line Flux Densities in MRS Channel 1\label{tbl:h2pah}}
\tablecolumns{8}
\tablewidth{0pt}
\tablehead{
\colhead{Location} &
\colhead{\molhy~S(5)} &
\colhead{\molhy~S(6)} &
\colhead{\molhy~S(7)} &
\colhead{PAH 6.2\um} & 
\colhead{\feii~$\lambda$5.34\um} & 
\colhead{\pfa} 
}
\startdata
E & 1.97 $\pm$ 0.05 & 0.42 $\pm$ 0.05 & 0.90 $\pm$ 0.08 & 83.82 $\pm$ 6.49 & 1.28 $\pm$ 0.06 & 0.25 $\pm$ 0.03 \\ 
N & 3.15 $\pm$ 0.09 & 0.67 $\pm$ 0.05 & 1.99 $\pm$ 0.12 & 93.40 $\pm$ 3.48 & 1.73 $\pm$ 0.09 & 0.28 $\pm$ 0.03 \\ 
NE & 1.76 $\pm$ 0.06 & 0.29 $\pm$ 0.06 & 0.81 $\pm$ 0.07 & 156.27 $\pm$ 4.84 & 1.95 $\pm$ 0.09 & 0.49 $\pm$ 0.03 \\ 
cen & 3.70 $\pm$ 0.13 & 0.82 $\pm$ 0.10 & 1.95 $\pm$ 1.09 & 121.82 $\pm$ 32.01 & 2.00 $\pm$ 0.14 & \dots $\pm$ \dots \\ 
NW & 1.26 $\pm$ 0.06 & 0.24 $\pm$ 0.02 & 0.68 $\pm$ 0.06 & 103.58 $\pm$ 5.36 & 1.49 $\pm$ 0.08 & 0.36 $\pm$ 0.03 \\ 
S & 4.88 $\pm$ 0.23 & 1.23 $\pm$ 0.19 & 2.50 $\pm$ 0.29 & 235.51 $\pm$ 7.74 & 4.21 $\pm$ 0.26 & 0.86 $\pm$ 0.15 \\ 
SE & 1.52 $\pm$ 0.09 & 0.27 $\pm$ 0.07 & 0.61 $\pm$ 0.08 & 173.99 $\pm$ 5.47 & 2.79 $\pm$ 0.13 & 0.66 $\pm$ 0.09 \\ 
SW & 4.46 $\pm$ 0.11 & 0.92 $\pm$ 0.14 & 2.30 $\pm$ 0.20 & 350.55 $\pm$ 7.97 & 5.73 $\pm$ 0.27 & 0.83 $\pm$ 0.11 \\ 
W & 7.89 $\pm$ 0.18 & 1.75 $\pm$ 0.11 & 4.56 $\pm$ 0.22 & 178.10 $\pm$ 9.80 & 3.77 $\pm$ 0.22 & 0.53 $\pm$ 0.08 \\ 
\enddata
\tablecomments{All flux densities are in units of 10$^{-23}$ W m$^{-2}$ pc$^{-2}$. The aperture used for the grid extraction is 0\farcs6 $\times$ 0\farcs6. \pfa~is not well detected at the center and thus its measurement is omitted.}
\end{deluxetable*}

\section{Discussion} \label{sec:discussion}

\subsection{Outflow Characteristics}
\label{sec:outflows}
\begin{figure*}[hbt!]
  \includegraphics[width=\textwidth]{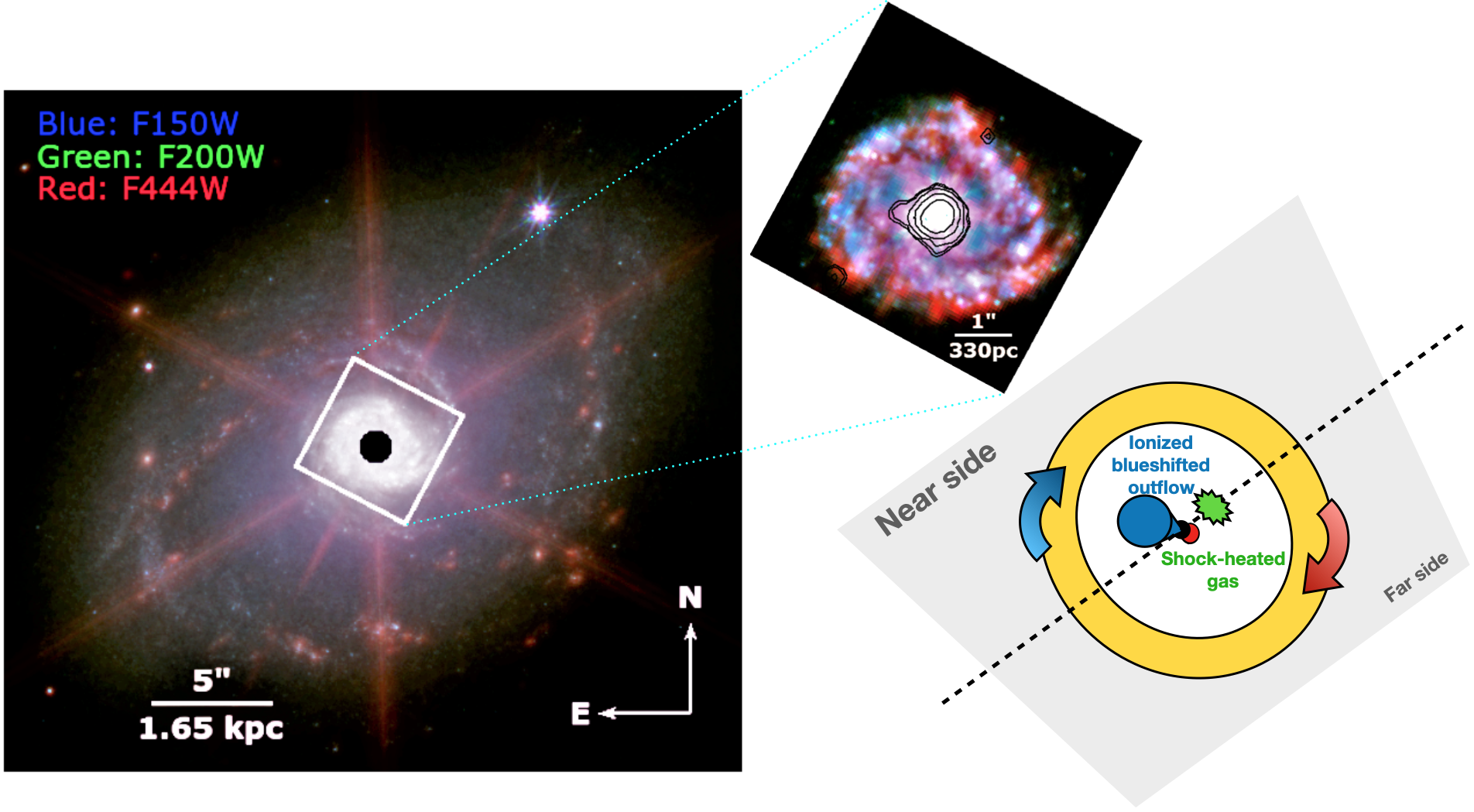}
  \caption{(Left) Three-color F150W/F200W/F444W NIRCam full-array image of NGC 7469 showing the large-scale spiral-arm structures. The inner core is saturated and thus masked out. 
  The overlaid box shows the subarray region. (Inset) Three-color F150W/F200W/F444W NIRCam subarray image showing the star-forming ring around the central nucleus. A 1\arcsec~(330 pc) scale bar is shown for comparison. The black contours indicate the intensity of the blueshifted component of the coronal line \mgv, most of which is coming toward us from the central AGN with a slight eastward extension. (Right) A cartoon schematic showing a nearly face-on outflow that appears one-sided and mostly in blueshift (blue cone). The inclined star-forming ring (\textrm{near side} in the N and \textrm{far side} in the S) 
  is rotating clockwise. Gas is shock-heated on the W side representing the interaction region where the outflow plows through the ISM of the galaxy. Components are not drawn to scale but merely to illustrate a plausible scenario based on the observed dynamics of the outflow, the inner ISM region, and the ring. 
  \label{fig:outflow}} 
\end{figure*}

As briefly introduced in Section \ref{sec:intro}, previous multiwavelength efforts in the literature converged on the presence of non-rotational kinematics in the circumnuclear region of NGC 7469, but the picture of how the non-rotating gas behaved at different physical scales was incomplete due to mismatches in observational parameters and data sensitivity~\citep{Muller-Sanchez11,Cazzoli20,Robleto-Orus21}. With \emph{JWST's} superb sensitivity, spectral resolution, and integral-field capability in the mid-infrared wavelengths, we can now incorporate the molecular, low- and high-ionization gases into a coherent picture with the same data set on the same physical scales.
We see from Figure \ref{fig:specs} that the high IP \mgv~line is prominently blueshifted, with a nearly face-on geometry and flux extending E of the nucleus. The intensity of the broad, blueshifted component from Figure \ref{fig:mgv} (lower left panel) is overplotted on the NIRCAM images as contours in Figure \ref{fig:outflow} (inset). 

Given that the gas is photoionized and \mgv~is tracing the illumination pattern, we see primarily the blue-shifted outflowing gas coming toward us from a nearly face-on galaxy with inclination angle of 45\textdegree~\citep{Davies04}, which is consistent with our measured $i$ of 51\textdegree~based on the ring structure. The \mgv~observations further support the picture in which the circumnuclear obscuration is roughly in the same plane as the galaxy's large-scale disk. 

To place our findings in the context of literature results, we consider the scales of the outflows detected with different instruments and tracers. 
\citet{Robleto-Orus21} found an outflow as characterized by blueshifted \oiii~$\lambda$5007\AA~extending 531 pc from the nucleus. 
Resolved at similar scales as our MIRI/MRS data, the \oiii~outflow may be more extended than the coronal wind we detect in \mgv~because it is more easily excited (with IP = 35 eV) and more extincted at the center. On the other hand, the \textrm{biconically-illuminated} \sivi~$\lambda$1.96\um~outflow presented by \citet{Muller-Sanchez11} has an E-W orientation, but at much smaller scale close to the AGN. 
The entire \sivi~emission (IP = 167 eV) falls within the central 1$-$2 spaxels of MIRI/MRS. The mid-infrared coronal outflow we find could have physical connection with the X-ray warm absorber in this AGN in terms of location and ionization condition~\citep{Blustin07}. Our results suggest a scenario where the high-ionization outflow detected by \emph{JWST} likely bridges the nuclear-scale ($<$ 100 pc) coronal line region outflow and the larger-scale narrow line region winds. The outflow appears one-sided because it is approaching us nearly face-on; the projected size of a potentially receding red-side cone behind the AGN may be too small to be spatially resolved by MIRI (Figure \ref{fig:outflow}). Even though there may be a hint of a redshifted wing in the W spectrum (Figure \ref{fig:specs}), the SNR is too weak in the individual spaxels for a robust spectral fit. The redshifted wind may simply be obscured by dense,  intervening ISM. 
Alternatively, the outflow may also appear asymmetric if an inclined jet is pushing on the gas in the disk~\cite[e.g. in NGC 1052;][]{Fernandez-Ontiveros19}. While NGC 7469 may host a \textrm{radio core--jet structure as resolved by the Very Large Array}~\citep{Orienti10}, no evidence of \textrm{radio} jet--ISM interaction has yet been found in its nuclear region~\citep{Xu22}.



\subsection{ISM excitation}
\label{sec:shocks}
Our MIRI/MRS observations allow us, for the first time, to examine the high-ionization outflow identified here in \mgv~at $r < 300$ pc in detail. 
We search for evidence that indicates how the outflow may impact its surroundings.
The mid-infrared rotational transitions of \molhy, when coupled with measurements of the PAH emission, provide a robust diagnostic of the excitation conditions and help determine the mechanism responsible for exciting the \molhy~lines~\citep{Higdon06,Ogle10,Guillard12,Cluver13}. Young massive stars may heat both PAHs and \molhy~in photodissociation regions (PDRs), but shocks arising from outflows or X-ray emission from AGN effectively dissociate PAH molecules and small grains~\citep{Jones96}. Empirically, low-luminosity star-forming galaxies exhibit a limited range of luminosity ratio $L(\mathrm{H}_2)/L(\mathrm{PAH})$ over several orders of magnitude in $L(\mathrm{H}_2)$ while those for AGN display line ratios $\sim$15 times greater~\citep{Roussel07}. Thus, the line ratio $L(\mathrm{H}_2)/L(\mathrm{PAH})$ has been used to distinguish between AGN, stars, and shocks as the driver of molecular emission in extreme environments such as LIRGs~\citep{Stierwalt14,U19}. 

Ideally, the total power in the warm \molhy~and all forms of PAH lines should be used in any analysis of PDR heating, given that the bulk of the molecular gas mass may be traced by the lower rotational \molhy~transitions. However, to take advantage of the high spatial resolution in MIRI Ch 1 and minimize contamination in the inner ISM region from the nucleus and the ring, we opt to confine our analysis of  $L(\mathrm{H}_2)/L(\mathrm{PAH})$ to the lines present in this channel, recognizing the potential limitations. In Ch 1 MRS data cube (Figure \ref{fig:specs}), four \molhy~transitions are identified: S(5) $\lambda$6.91\um~is detected and appears prominent in all the extracted spectra; S(6) $\lambda$6.11\um, while detected, sits on top of a prominent PAH feature at 6.2\um~and may be dominated by PAHs at certain locations; S(7) $\lambda$5.51\um~may be partially blended with \mgvii~$\lambda$5.5\um; and S(8) $\lambda$5.05\um~is comparatively weak. For these reasons, the S(5) line is the most optimal line among the possibilities. Its proximity to PAH6.2\um~reduces the risk of potential calibration differences between sub-bands and differential extinction, so we focus on using this pair of features for the $L(\mathrm{H}_2)/L(\mathrm{PAH})$ diagnostic. We note that close to the AGN ($<$ 300 pc), the \molhy~gas is expected to be quite hot~\citep[$\sim$ 900-1100K;][]{Pereira-Santaella22,Lambrides19} and is dominated by AGN heating in the central $\sim$100 pc (Armus et al. 2022, in preparation), so the S(5)~line likely carries much of the \molhy~luminosity, justifying our reliance on this line to measure the $L(\mathrm{H}_2)/L(\mathrm{PAH})$ ratio. 

\begin{figure*}[!hbt]
  \includegraphics[width=0.49\textwidth]{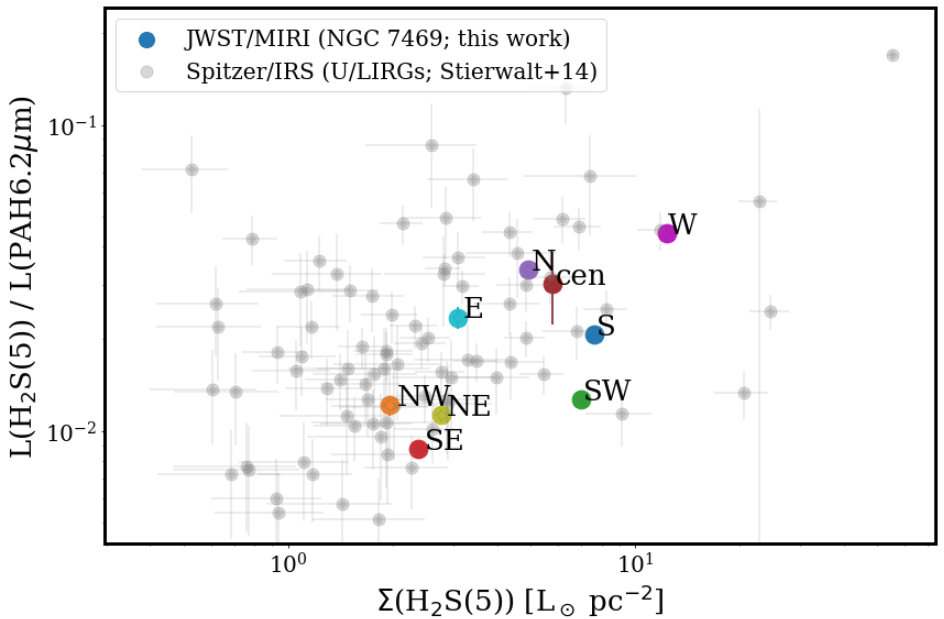}
  \includegraphics[width=0.49\textwidth]{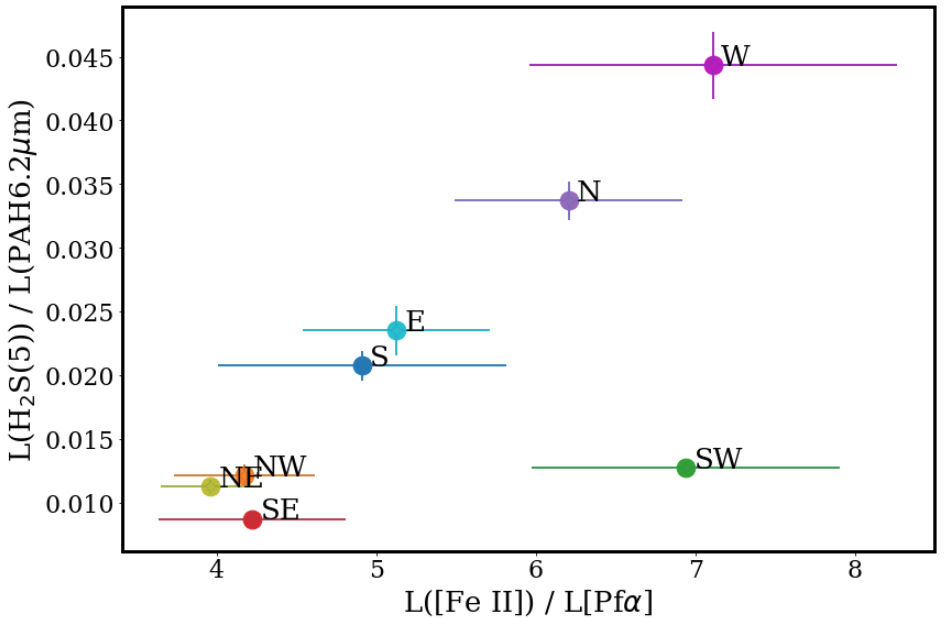}
  \caption{(Left) $L(\mathrm{H}_2)/L(\mathrm{PAH})$ vs. \molhy~luminosity density diagnostics diagram showing the data points corresponding to different parts of the inner ISM region. The grey points are local LIRGs from~\citet{Stierwalt14} for which these line transitions are detected and \molhy~luminosities normalized by the extraction aperture; upper limits have been excluded. 
The inner ISM in NGC 7469 tends to be bright in \molhy~but spans a range of $L(\mathrm{H}_2)/L(\mathrm{PAH})$ values nearly representative of the local LIRGs. (Right) The same $L(\mathrm{H}_2)/L(\mathrm{PAH})$ line ratios plotted against $L$(\feii)/$L$(\pfa). The correlation seen here ($\rho$ = 0.62$^{+0.26}_{-0.40}$ with a $p$-value of 0.10 for all the points; $\rho$ = 0.86$^{+0.09}_{-0.19}$, with a $p$-value of 0.01 excluding the SW outlier) suggests that shocks traced by \feii~are correlated with the shocked \molhy~gas. \pfa~is not well detected in the central spectrum and thus the center point is excluded from this plot. 
  \label{fig:h2pah}}
\end{figure*}

Figure \ref{fig:h2pah} (left) shows the $L(\mathrm{H}_2)/L(\mathrm{PAH})$ distribution versus \molhy~luminosity density for the different locations within the inner ISM. 
The first result is that the corner points (NW, NE, SE, and SW) all have lower $L(\mathrm{H}_2)/L(\mathrm{PAH})$~ratios than the other locations. Examining the spectra in Figure \ref{fig:specs},
the PAH features are more prominent at the corner locations that contain a larger fractional emission from the ring than at the E-W or N-S locations. This is expected, since PAH emission mostly originates from PDRs and, for some galaxy populations and/or environments, can be used as star formation rate tracers~\cite[e.g.][]{Peeters04}. 
In contrast, the \molhy~emission is relatively weaker in the ring, as seen in Figure \ref{fig:linemaps}, in agreement with what is expected from actively star-forming regions.

For comparison, we have plotted in Figure \ref{fig:h2pah} the integrated \emph{Spitzer}/IRS data points for a sample of local LIRGs~\citep{Stierwalt14} from the Great Observatories All-sky LIRGs Survey~\cite[GOALS;][]{Armus09}.  Because the angular-size aperture 
(3\farcs7 $\times$ 9\farcs5) 
used to extract the \emph{Spitzer}/IRS spectra 
is much larger than that of our extraction grid size, we show the \molhy~luminosity surface density, $\Sigma_{H_2}$, in the x-axis instead of luminosity, for a more direct comparison with the \emph{Spitzer}/IRS values. Our data points span a broad range in $L(\mathrm{H}_2)/L(\mathrm{PAH})$
of the GOALS LIRGs, and are consistent with the more \molhy~bright sources for a given $L(\mathrm{H}_2)/L(\mathrm{PAH})$ value. This shift may be due to the fact that the global values from \emph{Spitzer}/IRS are uniformly weighted and encompass 
many star-forming regions and obscured AGN sources,
while our resolved data points represent the inner ISM region close to the Seyfert nucleus in NGC 7469 where \molhy~is prominently detected.



The inner ISM region does not host bright star-forming regions and therefore PDR models are not applicable, but X-rays from the AGN and shocks from the outflow can heat the molecular gas~\citep{Petric18,Minsley20}.
Since iron is highly depleted onto grains, 
\feii~is a commonly adopted tracer of shock-excited gas where grains have been processed by outflows or other ionizing sources~\cite[see references in reviews by][]{Sajina22,U22}. Infrared studies of supernova remnants indicate that not only do they produce \molhy~lines as luminous as bright star-forming regions, they are also associated with enhanced \feii/\brg~relative to typical H \textsc{II} regions~\citep{Oliva89}. The latter observation has been predicted by theoretical models of shock fronts~\citep{McKee84}, where the ratio of \feii-to-Hydrogen is deemed a useful indicator of the efficiency of the shock front in destroying dust grains~\citep{Hollenbach84}. Here, we examine the ratio of $L$(\feii)/$L$(\pfa)~given that the H recombination line Pfund $\alpha$ traces star formation, provides a calibration baseline, and is detected in Ch 1.

Figure \ref{fig:h2pah} (right) shows $L(\mathrm{H}_2)/L(\mathrm{PAH})$ plotted as a function of $L$(\feii)/$L$(\pfa), which appears to be well correlated with the exception of the SW region. We compute a correlation for all the data points using \texttt{pymccorrelation}~\citep{Privon20} with 10$^4$ bootstrapping iterations to estimate the uncertainties. We determine the Pearson's correlation coefficient $\rho$ = 0.62$^{+0.26}_{-0.40}$, with a $p$-value of 0.10 driven mostly by the outlier point at SW and the large uncertainties of \pfa. If we exclude SW (which shows the peakiest PAH6.2\um~profile and the most extreme slope on the red end of the Ch1 spectra where \pfa~is located; see Figure \ref{fig:specs}), the resulting coefficient becomes $\rho$ = 0.86$^{+0.09}_{-0.19}$, with a $p$-value of 0.01, suggesting that \feii~and \molhy~at the W and N locations are likely excited by the same shocks driven by the outflow. This correlation between \feii~and \molhy~in (U)LIRGs has been studied by~\citet{Hill14}, who also inferred that shocks, including AGN-driven ones, are the likely source of \molhy~heating.

The most extreme region, W, which partially overlaps the region of high \molhy~and \feii~velocity dispersion (Figure \ref{fig:linemaps}), is, interestingly, on the opposite side of the center from the blueshifted coronal gas. 
It is possible that the \molhy~and \feii~are regions of high velocity dispersion represent regions where the receding, partially obscured part of the outflow is plowing into the dense ISM and releasing \feii~into the gas phase. A strong correlation between \feii~and \molhy~may be associated with sites of supernova remnants, where \feii~is produced in radiative shocks after grain destruction as the supernova remnant propagates~\citep{Hill14}. Given the extremely high-velocity coronal line components detected in the nuclear spectrum (Armus et al. 2022, in preparation), the AGN itself is likely driving the highly-ionized wind detected here in \mgv, which deposits energy into the ISM via shocks in the W and NW regions where the dispersion of the warm molecular gas and the $L(\mathrm{H}_2)/L(\mathrm{PAH})$ ratios are highest. It is somewhat surprising, given the location of the high dispersion \molhy~gas, that the NW spectrum is low in Figure \ref{fig:h2pah} (left).  However, this extraction region is large and partially intersects the inner edge of the ring, where the \molhy~is weak and the PAH is bright.  Examining the individual pixels in this region indicates the $L(\mathrm{H}_2)/L(\mathrm{PAH})$ ratio changes by a factor of eight as one moves from the upper right to lower left of this region.  Therefore the area of maximum dispersion in the NW would indeed be located much higher in Figure \ref{fig:h2pah}, close to the position of region W, consistent with our simple model.  A more detailed analysis of these regions will be performed in a subsequent paper using a spectral decomposition tool designed to produce accurate line ratio maps on finer scales.

\section{Summary} \label{sec:conclusions}
In this \emph{Letter}, we present new \emph{JWST} MIRI MRS observations of NGC 7469 and focus on the analysis of the inner ISM region between the central AGN and the starburst ring. The high spatial- and spectral-resolution available with the new IFS capability of MIRI enables a detailed investigation of the rotational \molhy, low- and  high-ionization fine structure lines, and dust features at mid-infrared wavelengths for the first time. We summarize our findings below. 

\begin{itemize}

    \item The morphology of the low-ionization \feii~and \arii~ lines are bright on the nucleus and in the starburst ring, showing several star-forming clumps and regions of enhanced emission.  The \molhy, in contrast, is strongly peaked on the nucleus and surrounding ISM, and is relatively weak in the starburst ring. 
    

    \item The \mgv~ emission line is resolved, and shows a broad, blueshifted component that is likely associated with the coronal line outflow in NGC 7469. The blueshifted highly-ionized outflow traced by \mgv, has a nearly face-on geometry that is strongly peaked on the nucleus, with an extension that reaches about 400 pc to the East. The \mgv~gas has a median line-of-sight velocity $v_\mathrm{med} = -205$ km s$^{-1}$ and reaches as high as $-650$ km s$^{-1}$ close to the AGN. 

    \item There are regions of enhanced velocity dispersion in \molhy~and \feii~$\sim$180 pc from the AGN that also feature excited $L$(\feii)/$L$(\pfa)~and $L(\mathrm{H}_2)/L(\mathrm{PAH})$~ratios, most clearly seen to the N and W. We identify these regions as the locations where the outflow is depositing energy into the dense interstellar gas via shocks, between the nucleus and the starburst ring.

    

\end{itemize}

Such a detailed view of the mid-infrared gas dynamics within the central region of a dusty LIRG nucleus is made possible for the first time, fully demonstrating the high spatial- and spectral-resolution plus high-sensitivity integral-field capability of \emph{JWST}.

\begin{acknowledgments}
We thank Jim Braatz, \textrm{Patrick Ogle, Andreea Petric, and the anonymous reviewer for suggestions that significantly improved the paper.}
This work is based on observations made with the NASA/ESA/CSA \emph{JWST}. The research was supported by NASA grant JWST-ERS-01328. The data were obtained from the Mikulski Archive for Space Telescopes at the Space Telescope Science Institute, which is operated by the Association of Universities for Research in Astronomy, Inc., under NASA contract NAS 5-03127 for JWST. These observations are associated with program \#1328 \textrm{ and can be accessed via \dataset[DOI:10.17909/0fe2-cf33]{http://dx.doi.org/10.17909/0fe2-cf33}}.
VU acknowledges funding support from NASA Astrophysics Data Analysis Program (ADAP) grant 80NSSC20K0450. 
The Flatiron Institute is supported by the Simons Foundation.
\textrm{HI and TB acknowledge support from JSPS KAKENHI Grant Number JP19K23462 and the Ito Foundation for Promotion of Science.}
AMM acknowledges support from the National Science Foundation under Grant No. 2009416.
ASE and SL acknowledge support from NASA grant HST-GO15472. YS was funded in part by the NSF through the Grote Reber Fellowship Program administered by Associated Universities, Inc./National Radio Astronomy Observatory.
SA gratefully acknowledges support from an ERC Advanced Grant 789410, from the Swedish Research Council and from the Knut and Alice Wallenberg (KAW) Foundation.
KI acknowledges support by the Spanish MCIN under grant PID2019-105510GB-C33/AEI/10.13039/501100011033.
F.M-S. acknowledges support from NASA through ADAP award 80NSSC19K1096.
Finally, this research has made use of the NASA/IPAC Extragalactic Database (NED) which is operated by the Jet Propulsion Laboratory, California Institute of Technology, under contract with the National Aeronautics and Space Administration.
\end{acknowledgments}

%

\facilities{\emph{JWST} (NIRCam and MIRI)}


\software{astropy~\citep{2013A&A...558A..33A,2018AJ....156..123A},
Cosmology calculator~\citep{Wright06},
\emph{JWST} Science Calibration Pipeline~\citep{jwstpipe},
ifscube~\citep{Ruschel-Dutra20,Ruschel-Dutra21},
BADASS~\citep{Sexton21},
CAFE~\cite[][D\`iaz-Santos et al. 2022, in preparation]{Marshall07}, 
JDAVis~\citep{Lim22},
PAfit~\citep{Krajnovic06},
pymccorelation~\citep{Curran14,Privon20}
          }



\appendix
\renewcommand\thefigure{\thesection.\arabic{figure}}    
\setcounter{figure}{0}    

\section{Fitting the Coronal Line}
\label{sec:mgv}

\begin{figure}[htb]
    \centering
    \includegraphics[width=.5\textwidth]{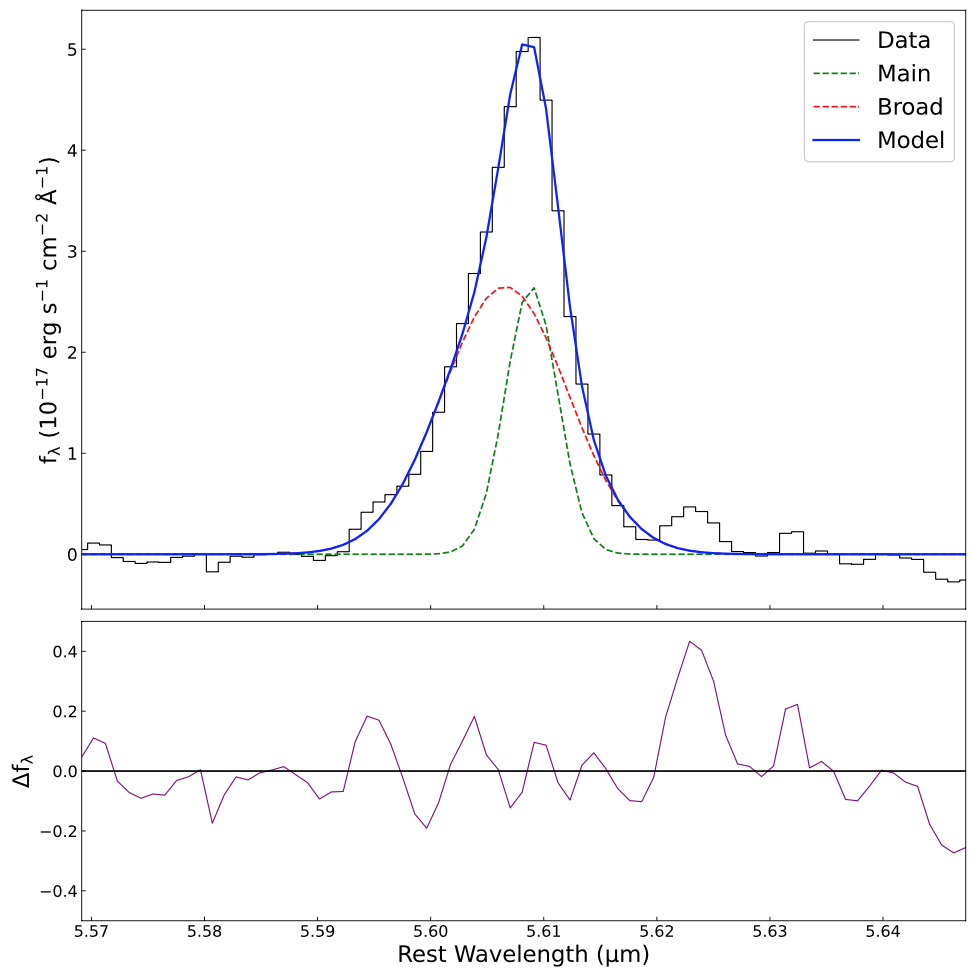}
    \caption{(Top) Example two-component Gaussian fit to the \mgv~from one spaxel at the center. 
    Two Gaussian components --- a main (green dashed) and a broad (red dashed) --- are needed to fit the ``outflow" region of the map. The residual resulting from subtracting Model (blue) from Data (black) is shown in the bottom plot. 
  \label{fig:badass}}
\end{figure}

\begin{figure}[htb]
    \centering
        \includegraphics[width=\textwidth]{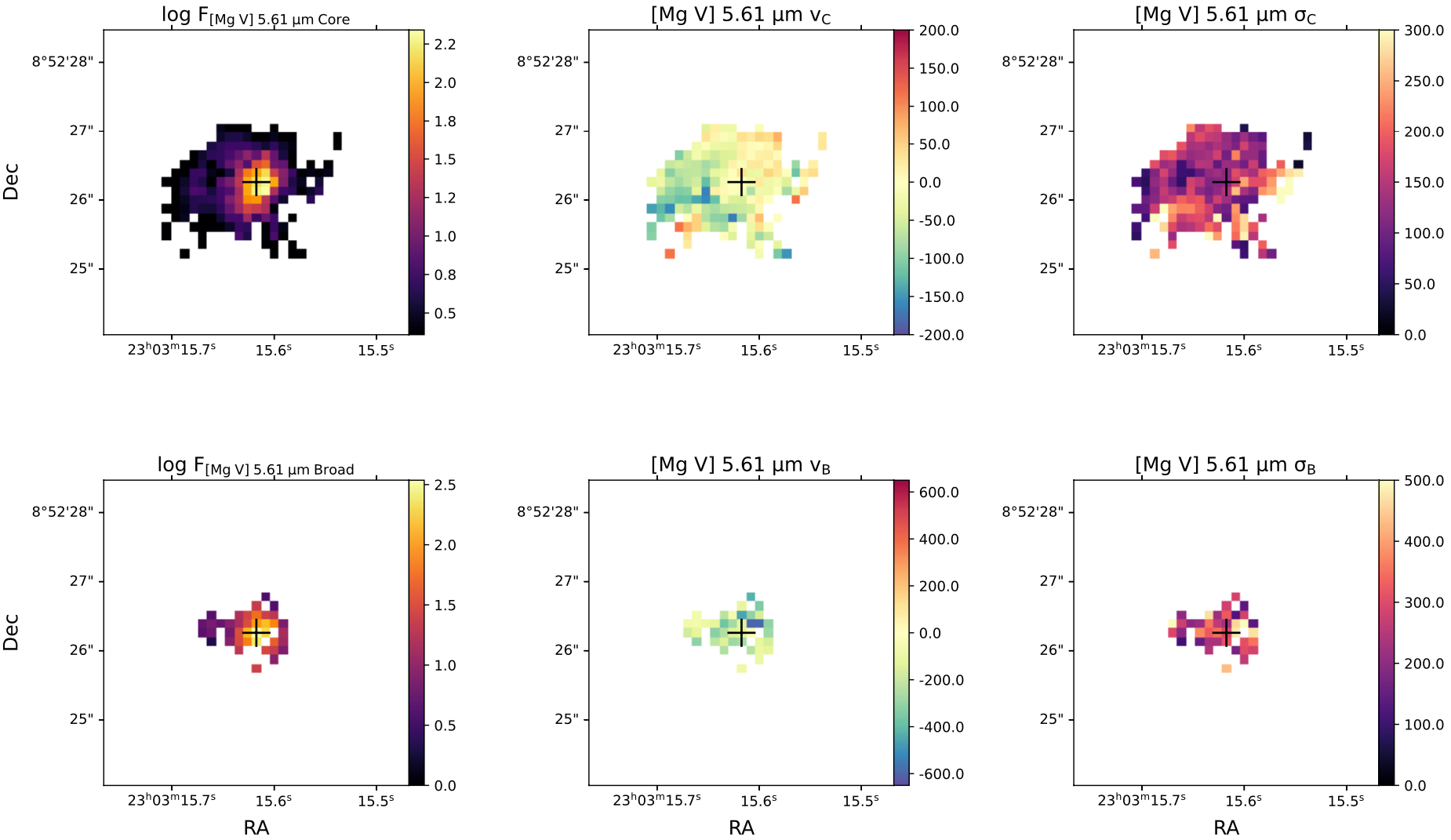}
    \caption{Flux (left, in log 10$^{-17}$ erg cm$^{-2}$ s$^{-1}$ px$^{-1}$), velocity (middle, in km s$^{-1}$), and velocity dispersion (right, in km s$^{-1}$) maps of the core, narrow Gaussian component (top) and of the second, broad, blueshifted component (bottom) of \mgv. The cross marks the location of the central AGN defined at the continuum. The velocity dispersion maps have been corrected for instrumental broadening. Note that the colorbar on these maps are all different in order to optimize visualizing the range of values spanned by each of the maps. North is up and East is to the left.
  \label{fig:mgv}}
\end{figure}

As mentioned in Section \ref{sec:data}, we conduct an in-depth investigation of the coronal line \mgv~over the entire Ch1 FOV using the BADASS~\citep{Sexton21} software package. We fit one- and two-component Gaussian profiles to the line spaxel-by-spaxel (e.g. Figure \ref{fig:badass}) and generate their resulting flux, velocity, and velocity dispersion maps (Figure \ref{fig:mgv}). 
The velocity dispersion maps have been corrected for instrumental broadening ($\sigma_\mathrm{inst} = 36.5$ km s$^{-1}$). The flux map of the secondary broad component of \mgv~(bottom left panel of Figure \ref{fig:mgv}) is overplotted on NIRCam images as contours in Figure \ref{fig:outflow} to indicate the high-ionization outflow.

We note that this two-component spectral fitting was only applied to \mgv~on a per-spaxel basis in this work to generate the moment maps. In a companion paper (Armus et al. 2022, in preparation), multi-component fitting is applied to all the emission lines in the full MIRI/MRS nuclear spectrum extracted from a small aperture on the AGN. 

\bibliography{ms_n7469}{}
\bibliographystyle{aasjournal}



\end{document}